\documentclass[11pt]{article}
\usepackage{amsmath,amssymb}
\usepackage{hhline}
\usepackage[scale={.75,.75}]{geometry}
\usepackage[T1]{fontenc}
\usepackage{bbm} 
\usepackage{fleqn} 
\usepackage{mathrsfs} 
\usepackage{cite} 
\usepackage{longtable}
\usepackage{verbatim}
\usepackage{graphicx}
\usepackage{upgreek}
\DeclareMathAlphabet{\mathpzc}{OT1}{pzc}{m}{it}
\usepackage[small]{caption} 
\usepackage{color}
\usepackage{psfrag}

\newcommand{\e}{\ensuremath{e}}

\newcommand{\be}{\begin{equation}}
\newcommand{\ee}{\end{equation}}
\newcommand{\bea}{\begin{array}}
\newcommand{\eea}{\end{array}}
\newcommand{\beq}{\begin{eqnarray}}
\newcommand{\eeq}{\end{eqnarray}}
\newcommand{\bes}{\begin{split}}
\newcommand{\ees}{\end{split}}
\newcommand{\pl}{\left\{}
\newcommand{\pr}{\right\}}
\newcommand{\al}{\left|}
\newcommand{\ar}{\right|}
\newcommand{\rr}{\right)}
\newcommand{\rl}{\left(}
\newcommand{\ccl}{\left[}
\newcommand{\ccr}{\right]}
\newcommand{\dd}{\partial}

\newcommand{\sigmab}{\bar \sigma}
\newcommand{\phib}{\bar \phi}

\newcommand{\Db}{\overline D}

\newcommand{\Wb}{\overline W}

\newcommand{\SU}[1]{\ensuremath{\mathrm{SU}(#1)}}

\DeclareMathOperator{\Tr}{Tr}

\numberwithin{equation}{section}
\numberwithin{table}{section}
\allowdisplaybreaks

\begin{document}

\date{}
\title{
\begin{flushright}
\normalsize{DESY-09-126}\\
\normalsize{NIKHEF/2009-017}\\
\end{flushright}
\vskip 2cm
{\bf\LARGE Dynamics of Moduli and Gaugino Condensates in an 
Expanding Universe}\\[0.8cm]}

\author{{\bf\normalsize
Chlo\'e~Papineau$^a$\!,
Marieke~Postma$^b$\! and
Sa\'ul~Ramos-S\'anchez$^a$}\\[1cm]
{\it\normalsize
${}^a$ Deutsches Elektronen-Synchrotron DESY, Hamburg, Germany}\\[2mm]
{\it\normalsize
${}^b$ NIKHEF, Kruislaan 409, 1098 Amsterdam, The Netherlands}
}

\maketitle

\thispagestyle{empty}

\vspace{1.3cm}

\begin{abstract}
\noindent
We study dynamical moduli stabilization driven by gaugino condensation in supergravity. In the presence
of background radiation, there exists a region of initial conditions leading to successful stabilization.
We point out that most of the allowed region corresponds to initial Hubble rate $H$ close to the scale of
condensation $\Lambda$, which is the natural cutoff of the effective theory.  We first show that
including the condensate dynamics sets a strong bound on the initial conditions. We then find that
(complete) decoupling of the condensate happens at $H$ about two orders of magnitude below $\Lambda$.
This bound implies that in the usual scenario with the condensate integrated out, only the vicinity of
the minimum leads to stabilization. Finally, we discuss the effects of thermal corrections.
\end{abstract}

\newpage\index{}

\section{Introduction}

Higher dimensional supersymmetric theories such as string theory are good candidates to explain the
origin of four-dimensional physics at low energies. In general, compactification of the additional
dimensions yields a large vacuum degeneracy, parametrized in terms of the so-called moduli fields. In
order to reproduce our observable universe and its particle content, consistent higher dimensional
theories must offer some mechanisms to fix the mass and the vacuum expectation value (vev) of the moduli.
In type IIB string compactifications, nontrivial background fluxes stabilize most of the geometrical
moduli~\cite{Giddings:2001yu}, but leave some other flat directions unlifted. In particular, the overall
volume parametrized by a K\"ahler modulus $\sigma$ is not fixed and hence requires the inclusion of
nonperturbative effects~\cite{Dine:1985rz,Banks:1994sg,Barreiro:1997rp}.

Gaugino condensation~\cite{Veneziano:1982ah,Taylor:1982bp} is perhaps the best understood and most
successful such mechanism capable of providing masses to moduli. It is a key ingredient in racetrack
models~\cite{Krasnikov:1987jj}, models with K\"ahler stabilization~\cite{Casas:1996zi}, and the
Kachru-Kallosh-Linde-Trivedi (KKLT) scenario~\cite{Kachru:2003aw}. The success of these models relies on
the assumption that the condensate forms at a high scale, inducing an effective scalar potential for
moduli. However, it is known that dynamical stabilization of moduli is fragile and depends on the
particulars of the universe evolution~\cite{Barreiro:1998aj}. If the cosmological expansion is dominated
by some energy source other than the moduli, such as radiation or vacuum energy, the initial conditions
leading to stabilization may be less restricted than in a modulus dominated universe. This statement has
been confirmed in matter and radiation dominated
scenarios~\cite{Brustein:2004jp,Barreiro:2005ua,vandeBruck:2007jw} including finite temperature
corrections~\cite{Barreiro:2007hb}.

Despite this progress, there is no reason to believe that only moduli dynamics will be affected by the
evolution of the universe. Indeed, it has been recently shown that in a universe undergoing fast
expansion, the condensate itself may be destabilized leading to unsuccessful moduli
stabilization~\cite{Lebedev:2009xu}. Thus, also the condensate is altered by the presence of additional
sources of energy. Whether gaugino condensation occurs depends on the dynamics of the gaugino pairs,
which can be described in terms of a complex (super)field $u$ in the Veneziano--Yankielowicz (VY)
approach~\cite{Veneziano:1982ah}. In the present work we take this view seriously and study the
consequences of the interaction between the condensate and background radiation upon moduli
stabilization. The main questions we will be concerned with are:
\begin{itemize}
\item when can the condensate field $u$ be safely integrated out without introducing inconsistencies in
  the theory?
\item for which initial conditions does gaugino condensation lead dynamically to moduli stabilization?
\end{itemize}
We shall show that for most configurations that lead to moduli stabilization, the condensate cannot be
integrated out and its dynamics must be considered.  Although, for illustrative purposes, we shall focus
on the KKLT model, it is straightforward to extend our findings to any model of moduli stabilization
driven by gaugino condensation, where similar results are expected.

\section{Single field dynamics}
\label{sec:dynamics}

As mentioned above, we consider the KKLT model~\cite{Kachru:2003aw} for moduli stabilization in the
context of type IIB string theory. After fixing the complex structure moduli and the dilaton, the
effective low energy ${\mathcal N} = 1$ four-dimensional supergravity theory only contains the volume
modulus $\sigma$, which is stabilized by gaugino condensation.\footnote{Note that a similar scenario
  arises in the weakly coupled heterotic string~\cite{Kappl:2008ie}, where $\sigma$ has to be replaced by
  the dilaton $S$.}  More specifically, an \SU{N} gauge theory localized on D7-branes wrapped around the
compactification volume has a field-dependent coupling\footnote{From now on, the subscripts $r$ and $i$
  denote respectively real and imaginary parts of the fields.}  $g^2 =1/ \sigma_r$.  If this sector is
asymptotically free, it undergoes gaugino condensation below a dynamical scale $\Lambda$. Using the VY
approach~\cite{Veneziano:1982ah}, it has been shown~\cite{Burgess:1995aa} that the only K\"ahler and
superpotential consistent with all symmetries are
\beq
\label{VY}
K &=& -3 \ln \rl \, \sigma +  \sigmab - \frac{u \bar u }{3} \, \rr \, , \\
W &=& u^3 \rl \, \sigma +  2 c \ln u \, \rr \, , \nonumber
\eeq
where the chiral superfield $u^3 \propto \Tr W^a W^a$ admits the condensate $\lambda^a \lambda^a$ as
lowest component. The constant $c = 3 N / \rl 16 \pi^2 \rr$ is the one-loop beta function coefficient.

In supergravity, the scalar potential is given by
\be 
V = \e^K \, \ccl \, K^{i \, \bar\jmath} D_i W \Db_{\, \bar\jmath}
\Wb \ - \ 3 \al W \ar^2 \, \ccr \quad , 
\label{Vsugra} 
\ee
where $K^{i \, \bar\jmath } = \rl K_{\bar\imath j}\rr^{-1}$ is the inverse K\"ahler metric, and $D_i W =
W_i + K_i W$ is the usual covariant derivative defined over the K\"ahler manifold. In the limit of small
$u \ll 1$, the dominant contribution to the potential is
\be
V_F \simeq \e^K |D_u W|^2 K^{u \bar{u}},
\label{approxVF}
\ee
which has as extrema $W = 0$, $W_u \simeq 0$ and $W_{uu} \simeq 0$, or
\begin{equation}
\label{eq:VYsolutions}
u = 0 \, , \qquad 
u = u_{\rm min} = \exp \rl -\frac{\sigma}{2c} -\frac{1}{3} \rr \, , \qquad 
u = u_{\rm max} = \exp \rl -\frac{\sigma}{2c} -\frac{5}{6} \rr \,.  
\end{equation}
The first solution $W =u=0$ formally corresponds to a supersymmetric chirally invariant vacuum. However,
the existence of such a vacuum is inconsistent~\cite{Cachazo:2002ry,Witten:1982df}. Arguably, this
minimum can be interpreted as an unstable nonsupersymmetric state~\cite{Shifman:1999mv}, where the VY
potential cannot be trusted. The other two solutions correspond to a minimum and a barrier separating it
from the unphysical state at the origin. Note that they are close in field space.  The minimum $u_{\rm
  min}$ is the VY-solution corresponding to gaugino condensation. Thus, we define the condensation scale
\be
\label{Lambda}
 \Lambda \equiv u_{\rm min}\;.
\ee
The mass of the field $u$ around the minimum is of order $\Lambda$, and therefore the field can be
integrated out at energies below this scale. Plugging the result into~\eqref{VY}, expanding in small $
u\bar u$ and adding a constant $W_0$ arising from integrating out the flux-stabilized moduli, we recover
the familiar KKLT potential
\beq
K &=&   - 3 \ln \rl \sigma + \sigmab \rr\, , \nonumber \\
W  &=&  W_0 \ + \ W_{\rm NP} \ = \ W_0 \ + \ A \, \e^{- a \sigma} \, , 
\label{KKLT}
\eeq
and identify $A = - 1/ \rl a e \rr$, $a = 3 / \rl 2 c \rr$. Note that for reasonable gauge groups, the
constant $a$ is bigger than one. Also, the inclusion of $W_0$ in~\eqref{VY} does not significantly
displace the position of the extrema~\eqref{eq:VYsolutions}.

Having clarified how gaugino condensation generates an effective potential for the modulus, we turn our
attention to the single field KKLT model.  As is well known, the scalar potential~\eqref{Vsugra} for
$\sigma$ exhibits a supersymmetric minimum
\beq
W_0 &=& - \ A \e^{-a \sigma_{\rm min}} \pl 1 \ 
+ \ \frac{a \rl \sigma_{\rm min} + \sigmab_{\rm min} \rr}{3} \pr 
\, , \label{sigmamin} \\
V_{\rm min}  &=&  - \ \frac{A^2 a^2 \e^{-a \rl \sigma_{\rm min} 
+ \sigmab_{\rm min} \rr}}{3 \rl \sigma_{\rm min} + \sigmab_{\rm min} \rr} 
\ < \ 0 \, , \nonumber
\eeq
of negative energy. We include a supersymmetry-breaking uplifting term
\be
\label{Vup}
V_{\rm up} \ = \ \frac{C}{X^p} \,,
\ee
where $X = \e^{-K/3}$ and $C$ may be tuned so as to obtain a Minkowski vacuum.  In the KKLT scenario the
uplift is provided by an anti D3-brane in a throat (in the bulk) of the compactification manifold, which
gives $p=2$ ($p=3$). The minimum~\eqref{sigmamin} is then separated from the runaway solution by a
barrier whose height is set by the gravitino mass.

Before discussing the dynamics of the system, let us comment on the approximations upon which we rely in
the remainder of the paper. It is known that (warped) string-inspired constructions, such as KKLT,
receive additional contributions, e.g. $\alpha'$ corrections~\cite{Becker:2002nn}, open-string loop
corrections~\cite{Berg:2005ja} and corrections due to (strong) warping~\cite{DeWolfe:2002nn,Chen:2009zi}.
Although they may significantly alter the form~\eqref{VY} of the supergravity action~\cite{Giddings:2005ff}, 
we do not incorporate these effects in our study. Our main concern is to understand whether a 
gaugino condensate can be integrated in/out when the dynamics of a modulus is taken into account. 
As such, we perform a comparative study between the single field behavior and the full system evolution. 
The corrections mentioned above are in this sense somewhat orthogonal to our purpose, and should not affect 
our conclusions. Moreover, let us emphasize again that we focus on the KKLT model as an illustrative example. 
Our observations hold for all moduli stabilization scenarios which rely on gaugino condensation, 
including for instance those without warping.

\subsection{Dynamical stabilization}
\label{subsec:review}

The barrier separating the KKLT minimum from the runaway solution at infinity is small compared to any
other natural scales in the problem. This is true, in particular, for the phenomenologically favored case
of low-energy supersymmetry breaking. At the same time, the scalar potential is exponentially steep. One
can therefore wonder how likely it is that the modulus will end up in the right minimum and not roll off
to infinity.  This is known as the overshoot problem, and has first been recognized
in~\cite{Brustein:1992nk}.  Dynamically, the higher the energy with which the field starts its dynamical
evolution, the more friction is needed in order to slow it down so that it finally settles in the
minimum.

In~\cite{Barreiro:1998aj,Brustein:2004jp,Barreiro:2005ua,vandeBruck:2007jw,Barreiro:2007hb}, it was shown
that the presence of a background fluid can help avoiding the overshoot problem by providing the
necessary friction. Indeed, there exists a relatively large range of initial conditions that lead to
successful stabilization. Clearly, this range grows with the initial background density, see e.g. Fig.~1
and 2 in~\cite{Barreiro:2007hb}. Throughout this paper, we consider the background fluid to be composed
of radiation. This allows us to work in a model-independent way. If the expansion of the universe were
instead driven by matter or vacuum energy density, the condensate would acquire a large mass of order
$H^2$ with a model dependent coefficient. Moreover, as was shown in~\cite{Lebedev:2009xu}, if the
background field gives the dominant contribution to SUSY breaking, the induced mass term destabilizes the
gaugino minimum at a scale $H \sim c \Lambda$, with $c \ll 1$, somewhat below the condensation scale~\eqref{Lambda}.
In a radiation dominated universe, the Hubble induced mass for the condensate and the volume modulus is small
and can be neglected at tree level~\cite{Lyth:2004nx}. For instance, a coupling of the form $u u^* T^{\mu}_{\mu}$, where $T$ is the stress-energy tensor of radiation, vanishes at tree level. One might however worry that this is not necessarily the case at one loop. We will come back to this point in Section \ref{sec:uandtwithT} when we consider finite temperature corrections.

In supergravity, the dynamics of a set of fields $\phi^i$ with non-canonical kinetic terms is described
by the equations of motion
\be
\label{eomsugra}
\ddot \phi^i \, + \, 3 H \dot \phi^i \, 
+ \, \Gamma^i_{j \, k} \dot \phi^j \dot \phi^k \, 
+ \, K^{i \, \bar\jmath} \dd_{\bar\jmath} V \ = \ 0 \, ,
\ee
where the Christoffel symbols are $\Gamma^i_{j \, k} = K^{i \, \bar l}
\frac{\dd K_{j \, \bar l}}{\dd \phi^k}$. The Hubble radius is subject
to the Friedmann constraint:
\be
\label{H}
3 H^2 \ = \  {\mathcal L}_{\rm kin} + V + \rho_\gamma  \quad ,
\ee
where ${\mathcal L}_{\rm kin} = K_{i \, \bar\jmath} \dot{\phi^i} \dot{\phib}^{\bar\jmath}$, and the
energy density of the background radiation satisfies
\be
\label{rhor}
\dot \rho_\gamma + 4 H \rho_\gamma \ = \ 0 \quad \Rightarrow \quad \rho_\gamma \ 
= \ \rho_{\gamma,\,{\rm ini}} \ \e^{-4 N} \, ,
\ee
where in the last equality, we have used as time variable the number of e-folds $N = \ln R$, with $R$ the
scale factor (recall that $H = \dot R / R$).  Here and in the following we use a subscript ${}_{\rm ini}$
to denote the corresponding quantity at the initial time $N=0$.  Without the background component, any
initial conditions satisfying $V(\sigma_{\rm ini}) > V(\sigma_{\rm max})$ will lead to overshoot of the
fields and consequently to a phenomenologically unacceptable runaway solution. However, as mentioned
before, if the energy density in the universe is dominated by radiation, the evolution of the scalar
fields is damped, and a much larger range of initial conditions leads to modulus stabilization. This
effect enters the equations of motion~\eqref{eomsugra} via a large Hubble rate $H \simeq
\sqrt{\rho_\gamma/3}$.

\begin{figure}[t!]
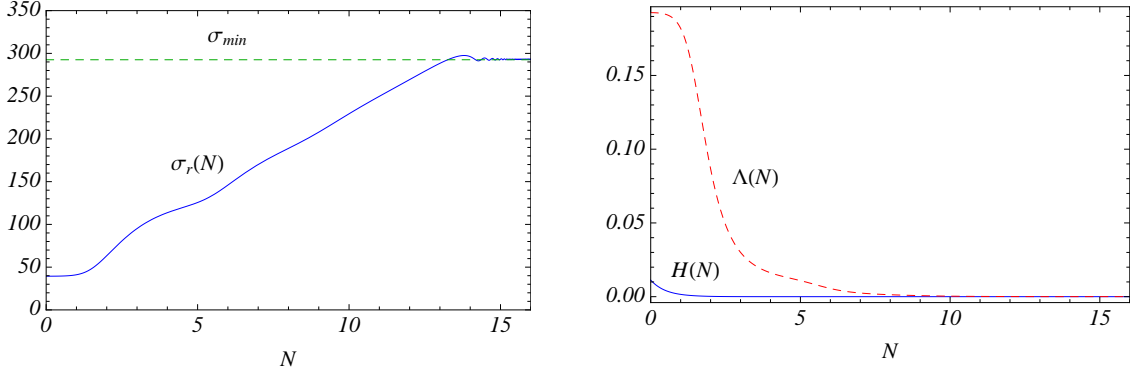

  \begin{minipage}{0.47\linewidth}
    {\centering
    \includegraphics[scale=0.7]{StabCase1.eps}
    }
  \end{minipage}
  \hspace{1mm}
  \begin{minipage}{0.47\linewidth}
    {
    \centering
    \includegraphics[scale=0.7]{StabCase2.eps}
    }
  \end{minipage}
  \caption{Example of initial condition leading to stabilization. The
  left panel shows the evolution of the field with time, in this case
  $\sigma_{r,\,{\rm ini}} \simeq 40$. The dashed green line gives the
  position of the minimum. On the right figure is plotted the
  evolution of $\Lambda$ (red, dashed line) and $H$ (blue, solid line)
  with time.}
  \label{fig:OneCase}
\end{figure}

The explicit equations of motion derived from~\eqref{eomsugra} for the KKLT model are given in
Appendix~\ref{app:eom}. Eqs.~\eqref{eommomenta} in the case of the potential~\eqref{KKLT} can be solved
numerically. With this purpose, we assume that the field has vanishing initial velocity
$\Pi_{\sigma,\,{\rm ini}} = 0$.  Following~\cite{Barreiro:2005ua,Barreiro:2007hb}, we parametrize the
initial radiation as
\be
\label{eq:rhogammai}
\rho_{\gamma,\,{\rm ini}} = \frac{\Omega_\gamma}{1 - \Omega_\gamma} \, V (\sigma_{r,\,{\rm ini}})\,, 
\ee
where we used the fractional energy density in radiation $\Omega_\gamma = \rho_{\gamma,{\rm
    ini}}/(3H^2)$.  Note that for $\Omega_\gamma \geq 1/2$ radiation initially dominates.  An example of
modulus stabilization is given in Fig.~\ref{fig:OneCase}, where we have taken $\Omega_\gamma = 0.99$. The
rest of the parameters is given as in~\cite{Kachru:2003aw,Barreiro:2005ua}, i.e $A=1$, $a=0.1$ and
$m_{3/2} \simeq 1$ TeV. The constant $W_0$ and $C$ are deduced from~\eqref{sigmamin} and~\eqref{Vup}.

\subsection{Consistency Condition}
\label{subsec:consistency}

As $\sigma_r$ rolls down the KKLT potential in the presence of the thermal fluid~\eqref{rhor}, the
corresponding scale $\Lambda = u_{\rm min} = \exp \ccl - \rl a \sigma_r + 1 \rr /3 \ccr$ evolves as well.
Indeed, only when the modulus is settled in its minimum does the gauge coupling of the hidden sector
stabilize as well. In order for the potential to follow effectively from~\eqref{KKLT}, we have to ensure
that condensation does occur even as the modulus is dynamically evolving. In other words, there is a
natural consistency condition: the Hubble rate $H$ has to be smaller than the condensation scale
$\Lambda$ {\it at all times}.

It can be seen from the right plot of Fig.~\ref{fig:OneCase} that the scale $\Lambda$ for this specific
example is indeed higher than $H$ and remains as such with time. Actually, because the field starts with
vanishing velocity, it is clear that if $H_{\rm ini} \lesssim \Lambda_{\rm ini}$, then the same holds at
later times.

\begin{figure}[t!]
  \centering
    \includegraphics[scale=0.7]{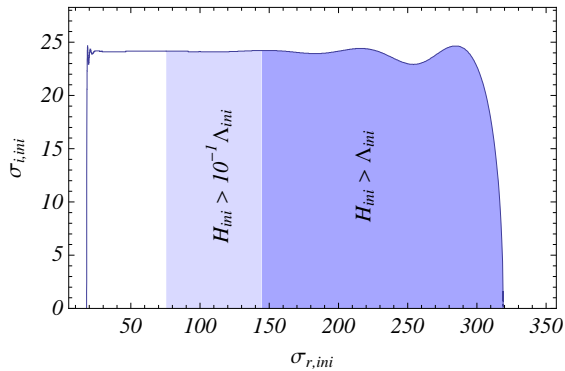}
    \caption{Set of initial conditions $\{\sigma_{r,\,{\rm
      ini}},\,\sigma_{i,\,{\rm ini}}\}$ leading to stabilization at the
      minimum of the KKLT potential for $\rho_{\gamma,\,{\rm
      ini}}=10^{-4}$, $A=1,\, a=0.1$ and $m_{3/2} = 1$ TeV. The solid
      line corresponds to the result of~\cite{Barreiro:2007hb}.  The
      darker region can be excluded since $H_{\rm ini} > \Lambda_{\rm
      ini}$.}
\label{fig:StableInitialConditions}
\end{figure}

In Fig.~\ref{fig:StableInitialConditions} we show how the region of successful initial conditions for
$\sigma$ found in~\cite{Barreiro:2005ua} is modified once we require that these points are consistent
with having the condensate initially formed. Actually, even if the condensate is initially formed, and
thus remains so at later times, it is legitimate to wonder below which scale it can be safely integrated
out. This is why we also show the bound $H_{\rm ini} \leq 10^{-1}\Lambda_{\rm ini}$ in
Fig.~\ref{fig:StableInitialConditions}. Formally, a field can be integrated out once its oscillations
around its minimum are negligible. Hence, for a large range of initial parameters, consistency requires
that we integrate the condensate in, and consider the whole system~\eqref{VY} together with $W_0$. In the
next section, we let the two fields evolve and study the region of stabilization, putting particular
emphasis on the way the condensate may be integrated back out.

\section{Dynamics of the gaugino-modulus system}
\label{sec:full}

In this section we discuss the dynamics of the K\"ahler modulus with the gaugino condensate integrated
in. For simplicity, in what follows, we concentrate on the real parts of $\sigma$ and $u$, and omit
subscripts.  Dynamics of the imaginary parts shall be treated elsewhere~\cite{Papineau:2009xx}.

The VY effective potential for gaugino condensate $u$ is given in~\eqref{VY}, the additional uplift term
in~\eqref{Vup}.  As the Hubble scale approaches the condensation scale, the evolution of the full
condensate-modulus system starts to deviate from the KKLT description, which is only a good approximation
for small $u$ when~\eqref{approxVF} is valid.  This can be seen explicitly from the F-term potential
\be
\label{VFfull}
V_F = \frac1{X^2} \ccl -2 W_0 u^3 + (2c+3 \sigma +6c \ln u)^2 u^4
  +\frac{1}{3}(4c + 2 \sigma)u^6 \ccr
\ee
with $X = \e^{-K/3}$, as before.  The quartic term in $V_F$ is proportional to $|W_u|^2$. In the limit
that $u \ll 1$ and $W_0$ small, both the cubic and the $u^6$ term are subdominant, and we recover the VY
result that $W_u=0$ minimizes the potential. The scalar potential is a surface; it exhibits a valley of
attraction corresponding to the direction $u_{\rm min} (\sigma)$. This valley is much steeper in the
condensate direction than along the modulus direction.  Along the $u$ direction there is a barrier
separating the ``good'' minimum, from the state at $u=0$; in the modulus direction there is the usual
barrier separating the ``good'' KKLT minimum from the runaway solution at infinity.

The equations of motion for $\sigma$ and $u$ momenta in a spatially flat Friedmann-Robertson-Walker
spacetime are given by~\eqref{eomsigmaandu}. As before, we assume an homogeneous background radiation
energy density, which evolves in an expanding universe according to~\eqref{rhor}.

We integrated numerically the equations of motion for the real fields $\{\sigma, u\}$ keeping the phases
fixed at their instantaneous minimum.  The free parameters in the model are $c$, $W_0$ (which determines
the vacuum gravitino mass), and $\Omega_{\gamma}$. In what follows, we choose $c=1/7$ and a) $W_0 = 10^{-6}$ and b) $W_0 =
10^{-8}$. For these parameters, the minimum with the modulus stabilized is at $\sigma_{\rm min} =
1.22 \,(1.68)$ and $u_0 = u_{\rm min}(\sigma_{\rm min}) = 0.01 \; (0.002)$ for $W_0 = 10^{-6}
\,(10^{-8})$.  Demanding a Minkowski vacuum, we obtain the uplifting parameter $C\sim3\cdot10^{-13}\;
(3\cdot10^{-17})$ for $W_0 = 10^{-6} \,(10^{-8})$. We used relatively large values of $W_0$, giving rise
to high scale supersymmetry breaking, because of numerical convenience.  Nevertheless, we expect that our
qualitative results can be straightforwardly extrapolated to lower scales.

In our study, we take a conservative approach and assume that whenever $u$ jumps the barrier at $u_{\rm
  max}$ and is driven towards the state at the origin, the gauginos do not condense and, consequently,
$\sigma$ does not get stabilized.  Apart from an understanding of the physics at the minimum $u=0$,
discussing the dynamics of $u$ in this case requires to include quantum effects, such as particle
production of fields which are light at the special symmetry point at the origin~\cite{Kofman:2004yc}.
Particle production damp the motion of $u$, possibly trapping it in the region $|u| < u_{\rm max}$ and
leaving our results untouched.  Another possibility is that the field has enough momentum to go through
the origin and bounce back over the barrier to end up at $u_{\rm min}$.  This could open up an additional
region of preferable initial conditions, which nonetheless will strongly depend on the details of the
potential near the origin.

\subsection{When can the gaugino condensate safely be integrated out?}

The first issue to address is to determine for which initial energy scales $H_{\rm ini}$ it is a good
approximation to integrate out the gaugino condensate and work in the low energy effective KKLT theory,
and for which initial conditions the dynamics of the full system should be taken into account.  To answer
this question we determined the range of initial conditions for $\sigma_{\rm ini}$ leading to modulus
stabilization in the effective KKLT model, with $u$ integrated out.  We compared the results with the
range of $\sigma_{\rm ini}$ leading to stabilization in the full system, where both $u$ and $\sigma$ are
kept as dynamical fields, starting with $u$ initially lying in the valley of attraction, i.e. $u_{\rm
  ini} = u_{\rm min}(\sigma_{\rm ini})$. We numerically determined this $u_{\rm ini}$, as the analytic
approximation~\eqref{eq:VYsolutions} breaks down for small $\sigma_{\rm ini}$ (which corresponds to large
$u$). In both cases we started with zero momenta for all fields.

\begin{figure}[t!]
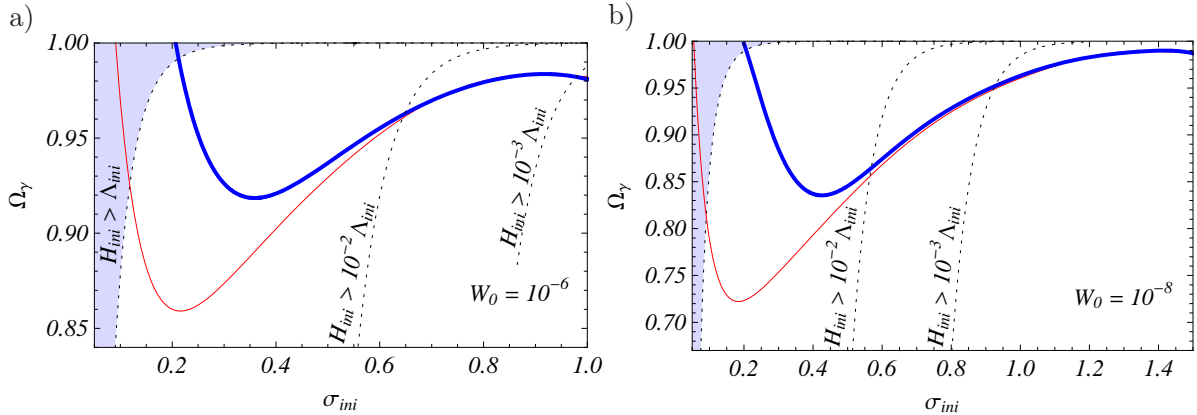

  \begin{minipage}{0.47\linewidth}
    a)\\
    {\centering
    \includegraphics[scale=0.75]{plot10-6.eps}
    }
  \end{minipage}
  \hspace{1mm}
  \begin{minipage}{0.47\linewidth}
    b)\\
    {
    \centering
    \includegraphics[scale=0.74]{plot10-8.eps}    
    }
  \end{minipage}
  \caption{The regions bounded by the curves correspond to initial conditions leading to modulus
    stabilization in the KKLT model with $c=1/7$ and a) $W_0=10^{-6}$ and b) $W_0=10^{-8}$ as a function
    of the initial abundance $\Omega_\gamma$. The thin/red curve is obtained in absence of $u$ and the
    thick/blue curve is for $u$ integrated in and initially at its minimum $u_{\rm min}$. Black dotted
    lines correspond to $(H/\Lambda)_{\rm ini}=1,10^{-2},10^{-3}$.}
\label{fig:comparison}
\end{figure}

The space of initial conditions $\{\sigma_{\rm ini},\,\Omega_{\gamma} \}$ for which the system evolves to the overall minimum is plotted in
Fig.~\ref{fig:comparison}; for all other initial conditions, $u$ and/or $\sigma$ overshoot. The thin/red
curve gives the stabilization region for the KKLT potential with $u$ integrated out.  The results agree
with previous findings~\cite{Barreiro:2005ua}: provided that the initial energy density is dominated by
radiation, there is a large range of initial $\sigma_{\rm ini}$ leading to stabilization.  Starting with
initial values $\sigma_{\rm ini}$ to the left of the curve, the modulus picks up too much momentum and
overshoots; starting with initial values $\sigma_{\rm ini}$ to the right of the curve, the modulus is too
close to the minimum and there is not enough time to damp the field, with overshoot as result.
Finally, we note that there is another small stabilization region for $\sigma_{\rm ini}$ close to the
minimum $\sigma_{\rm min}$ (that is not displayed in the plot), for which the initial potential energy is
less than the height of the barrier $V_{\rm ini} < V_{\rm max}$.

The equivalent curve for the full system is the thick/blue line. As expected, there are large
discrepancies for small $\sigma_{\rm ini}$, and thus large $u_{\rm min}(\sigma_{\rm ini})$, whereas for
larger $\sigma_{\rm ini}$ both systems give the same results.  In terms of $n =(H/\Lambda)_{\rm ini}$,
the deviations are negligible when this ratio drops below $n = 10^{-2}$, and become large around $n =
10^{-1}$ --- Fig.~\ref{fig:comparison} shows the lines for $n=1,\,10^{-2},\,10^{-3}$.  This is almost
independent of $W_0$, i.e. of the vacuum gravitino mass; the reason is that for such initial conditions
the cubic term in the potential~\eqref{VFfull} is negligible, and all $W_0$ dependence drops out. This
allows to restate the results in terms of $V_{\rm ini}$ as well: for $V_{\rm ini} \lesssim 10^{-7}$ the
KKLT and full potential behave very similarly, whereas for $V_{\rm ini} \gtrsim 10^{-5}$ they deviate
significantly.

We note that the smaller $W_0$ is, the longer it takes for the fields to roll down to their minima, and 
the longer they are exposed to damping; this explains why less initial radiation $\Omega_{\gamma}$ is 
needed for stabilization. Hence, for a TeV gravitino mass, $W_0 \sim 10^{-14}$ and $c = 1/10$, it is 
natural to expect the same behavior as in Fig.~\ref{fig:comparison}, with an even larger accessible 
range of $\Omega_\gamma$. In that case, one can further show that stabilization occurs at the 
unification-compatible point $g^2 = 1/2$.

As already stressed, the main reason for the breakdown of the KKLT approximation at large initial energy
densities is due to the way the condensate is integrated out. Formally, the field $u$ should be
integrated out at the level of the scalar potential. In the KKLT approach, however, this is not the case:
the field $u$ is sent to zero in the K\"ahler potential, and replaced by its vev in the superpotential.
The difference between the two approaches is easy to understand by studying the structure of the scalar
potential~\eqref{Vsugra}; it gives rise to extra terms when the first approach is followed. The way heavy
fields can be integrated out consistently in supergravity has been the center of renewed attention in the
past months~\cite{Achucarro:2008sy,Gallego:2008qi,Brizi:2009nn}. In this paper, we are dealing with
intermediate energy ranges for which the extra terms in the scalar potential are not negligible. They
have an impact on the dynamics of the system. Therefore, for small $\sigma_{\rm ini}$, the solution for
$u_{\rm min}$~\eqref{eq:VYsolutions} is no longer a good approximation, and thus higher order terms in
$u$ (the $u^6$-term in eq.~\eqref{VFfull}) have to be taken into account. Numerically, we find that the
exact and the analytical solutions start to deviate more and more as we increase $V_{\rm ini}$.  Notice
that for $\sigma_{\rm ini} < 0.2$ (or $V_{\rm ini} \gtrsim 10^{-3}$), the nontrivial minimum of
$V_F(\sigma=\sigma_{\rm ini},\,u)$ along the $u$ direction disappears altogether, implying that gaugino
condensation cannot occur.  This explains the cutoff in the stabilization region in both plots. For
$\sigma_{\rm ini} < 0.2$, any $u_{\rm ini}$ leads to overshoot.

\subsection{Initial conditions leading to modulus stabilization}

We now determine the stabilization region in $\{\sigma_{\rm ini}, u_{\rm ini}\}$ space, considering
initial conditions with both the modulus and the condensate displaced from the instantaneous minimum.
Once again the fields start their evolution at rest.

\begin{figure}[tbp]
\hspace{-1cm}
\centering
\includegraphics[scale=0.8]{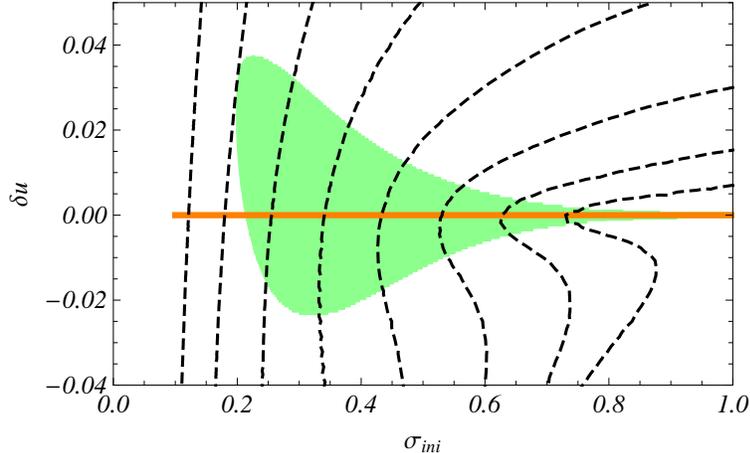}
\caption{Initial conditions $\{ \sigma_{\rm ini},u_{\rm ini} - u_{\rm
  min}(\sigma_{\rm ini}) \}$ leading to modulus stabilization for $W_0
  = 10^{-6}$. The black dashed lines correspond respectively to
  $V_{\rm ini} = 10^{-1},10^{-2},\dots,10^{-8}$ from left to right.}
\label{fig:scan}
\end{figure}

The results for $W_0 = 10^{-6}$ and $\Omega_{\gamma} = 0.95$ are shown in Fig.~\ref{fig:scan}.  The
initial conditions $\{\sigma_{\rm ini},\,\delta u \equiv u_{\rm ini} - u_{\rm min}(\sigma_{\rm ini}) \}$
leading to stabilization lie in the shaded/green area.  A large range of modulus initial values leads to
stabilization $0.2 \lesssim \sigma_{\rm ini} \lesssim 0.6$, whereas only a small displacement in $\delta
u$ is allowed. The reason for this is simple: the potential is much steeper in the condensate direction,
even a comparatively small displacement $|\delta u| \lesssim 2 \times 10^{-2}$ implies a large increase
in $V_{\rm ini}$, and consequently leads to overshoot.

To address the likelihood of initial conditions leading to stabilization, it is therefore useful to
consider not the initial field displacement, but rather the corresponding change in energy $V_{\rm ini}$.
Given an initial energy $V_{\rm ini}$, and random initial conditions for both the modulus and condensate
(but such that $V_{\rm ini}$ remains fixed), how large is the stabilization region in the space of
initial conditions?  In Fig.~\ref{fig:scan} are plotted the equipotential lines corresponding to $V_{\rm
  ini} = 10^{-1},10^{-2},\dots,10^{-8}$.  The lines are approximately orthogonal to $\delta u~=~0$ for
small $\sigma_{\rm ini}$.  Considering all initial conditions along one such line to be equally likely,
the stabilization region is very small.  Most initial conditions lead to overshoot. This is quite a
different result than the conclusion one (naively) draws from the one-dimensional system with the
condensate integrated out.  Confining to initial conditions to the left of the maximum, a good proportion
of initial conditions leads to stabilization.

The solid/orange line at $\delta u=0$ in Fig.~\ref{fig:scan} corresponds to a similar study in the
effective KKLT model, without the dynamics of $u$. The points along that line at the left of the
shaded/green area are initial conditions that appear to lead to stabilization from the one-dimensional
standpoint. Once the dynamics of the condensate are considered, it is clear that those points are
inconsistent.

\section{Finite temperature corrections}
\label{sec:uandtwithT}

In this section we study the effect of thermal corrections on the evolution of the modulus-condensate
system. These corrections arise because the gauge kinetic term $W^{a} W^{a}$ of the thermal sector
acquires an expectation value at one loop~\cite{Grisaru:1983hc,Buchmuller:2003is}, and because the
particles in the thermal bath may have a field dependent gauge or Yukawa coupling.  As an example, let
us consider that the background thermal fluid $\rho_\gamma$ originates from an \SU{N_c} gauge sector with
$N_f$ matter multiplets (in the fundamental representation) and a modulus-dependent gauge coupling, $g^2
= 1 / \sigma$. Even though the (real) modulus $\sigma$ does not thermalize, its potential receives
corrections through the gauge coupling of the thermal bath. Hence, at a constant temperature $T$, the
effective potential energy of the modulus is
\begin{equation}
\label{eq:effectiveV}
V(\sigma) + F (g,\, T)\,,
\end{equation} 
where $V(\sigma)$ is the zero-temperature potential~\eqref{VFfull} and the induced free energy is given
by\footnote{Formally, eq.~\eqref{eq:freeenergy} is valid only in the weak coupling regime. If
  $g\sim\mathcal{O}(1)$, additional corrections become important. However, explicit computations have
  shown that the $g^2$ term contains already the correct qualitative behavior~\cite{Kajantie:2002wa}.}
\be
\label{eq:freeenergy}
F \rl g,\, T \rr \ = \ \rl a_0 + a_2 g^2 \rr \, T^4 \,, 
\ee
where $a_0 = - \frac{\pi^2}{24} \rl N_c^2 + 2 N_c N_f - 1 \rr$ and $a_2 = \frac{1}{64} \rl N_c^2 -1 \rr
\rl N_c + 3 N_f \rr$.  Note that for a sufficiently high temperature $T_{\rm crit}$, the minimum of the
potential gets lifted and, if the modulus is already in its minimum, it gets
destabilized~\cite{Buchmuller:2004xr,Buchmuller:2004tz}. 

Similarly, since the trace of the stress-energy tensor no longer vanishes, a mass term for the condensate
may be generated\footnote{We thank Oleg Lebedev for drawing our attention to this issue.} via a coupling
$u u^* \,\langle T^{\mu}_{\phantom{.}\mu}\rangle_{\phantom{.}_T}$. It is however easy to show that the
corresponding destabilization temperature is (much) higher than the one affecting the modulus, 
$T_{\rm crit}$. We have confirmed numerically that such a mass term can be neglected.

The existence of a thermal bath affects the moduli dynamics.  First, the solution of Friedmann's
equations for the radiation density becomes
\be
\label{eq:rhowithT}
\rho_\gamma ~=~\rho_\gamma^{\rm init}\e^{- 4 N} 
\rl\frac{1 + r g^2\rl \sigma_{\rm init} \rr}{1 + r g^2\rl
  \sigma \rr} \rr^{1/3} \,, 
\ee 
where $r$ denotes the ratio\footnote{Our definition of $r$ differs by a factor $4\pi$ compared
  to~\cite{Barreiro:2007hb}. This originates from the different dependence of the gauge coupling on
  $\sigma$.}  $a_2 / a_0$ and $N = \ln R$ as before.  Since $\rho_\gamma$ enters the Hubble rate $H$,
eq.~\eqref{eq:rhowithT} implies that temperature corrections introduce an additional source of friction
which decreases as the universe expands. As a result, the equations of motion~\eqref{eomsugra} are
modified as follows
\begin{equation}
  \label{eq:VtoVeff}
\dd_{\bar\jmath} V \to \dd_{\bar\jmath}   (V+F)
 ~=~ \dd_{\bar\jmath}  V - \frac13 \frac{r\rho_\gamma}{1+rg^2} \dd_{\bar\jmath} g^2\,,    
\end{equation}
where we have used the relation between the energy density and the temperature of the thermal fluid
$\rho_\gamma~=-p_\gamma+T dp_\gamma/dT ~=-3a_0(1+r g^2)\,T^4$ \ with the pressure given by
$p_\gamma=-F(g,\,T)$.  This new contribution to the equations of motion renders manifest the effect of
the effective potential~\eqref{eq:effectiveV}. Note that irrespective of whether the initial temperature
is larger than the temperature at which the minimum of the potential gets lifted, $T_{\rm ini}>T_{\rm
  crit}$, the modulus can still be stabilized. If the friction is large enough, the minimum emerges
before the modulus has had time to approach it. Some of the initial conditions that lead to stabilization
at zero temperature are however erased. Thus, the region of admissible conditions reduces under the
influence of the thermal bath.

We note that it is possible that a subset of the thermal degrees of freedom do not have a
modulus dependent gauge or Yukawa coupling, and only interact gravitationally with the modulus sector.
If this subset is large, the parameter $|r|$ can be made arbitrarily small, and the results of the previous
section apply.

\subsection{Initial conditions leading to modulus stabilization}

\begin{figure}[t!]
 \centerline{\includegraphics[scale=0.75]{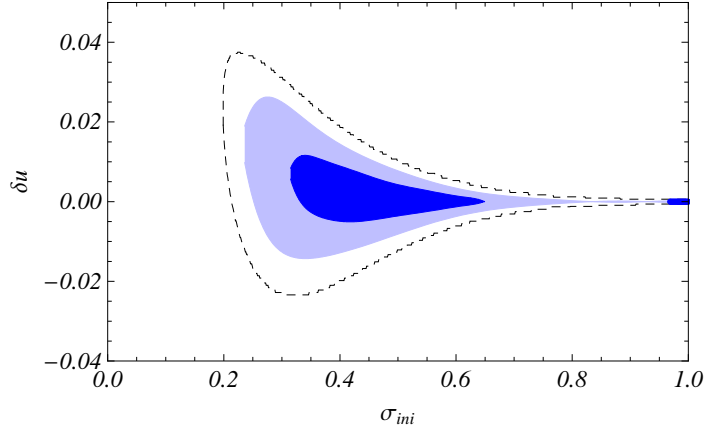} }
 \caption{Region of initial conditions leading to $\sigma$ stabilization in the KKLT model.  The dashed
   line surrounds all admissible initial values in the absence of thermal corrections.  The light-shaded
   (dark-shaded) region corresponds to the allowed parameters in presence of thermal corrections, with
   $r=-0.04$ ($r=-3/4\pi^2$). The chosen parameters are $c=1/7$, $W_0=10^{-6}$, $\Omega_\gamma=0.99$.}
 \label{fig:TplottVsdu}
\end{figure}

As before, we consider the KKLT model with the condensate field $u$ integrated in and the parameters
$c=1/7$ and $W_0=10^{-6}$. In addition, we take the initial radiation abundance to be
$\Omega_\gamma=0.99$ and consider two different configurations with $r=-3/4\pi^2$ and $r=-0.04$ to illustrate the dependence on thermal corrections. The first case corresponds to an \SU2 pure Yang-Mills thermal sector and is thus the minimal $\al r \ar$ achievable when the contribution~\eqref{eq:freeenergy} is the only component in radiation. Fig.~\ref{fig:TplottVsdu} shows the set of admissible initial conditions $\{\sigma_{\rm ini},\,\delta
u\}$ that lead to stabilization.  The region surrounded by the dashed line corresponds to the result
obtained in the previous section for zero temperature (cf. Fig.~\ref{fig:scan}). Inside this region, we
find those admissible initial conditions for $r=-0.04$ (light shaded) and $r=-3/4\pi^2 \approx - 0.076$ (dark shaded).

A few comments are in order. The region of stabilization clearly reduces when $\al r \ar$ increases. For the SM gauge group \SU3$\times$\SU2, one finds $r \sim -0.1$, which leads to a considerable reduction of the allowed region. This implies that only for an even more fine-tuned value $\Omega_\gamma \sim 1$ does stabilization occur in a more realistic scenario. Also, the smallest value that the modulus can initially take depends on
the thermal background. This happens because the higher the potential energy of the modulus $V_{\rm ini}$
is, the faster it rolls down to the position of the minimum. If the temperature is not sufficiently low,
the modulus finds no minimum and consequently runs away. Similarly, if $\sigma$ starts too close to the
local maximum of the modulus potential, at higher temperatures the modulus does not feel the attraction
of the minimum and escapes.

\begin{figure}[t!]
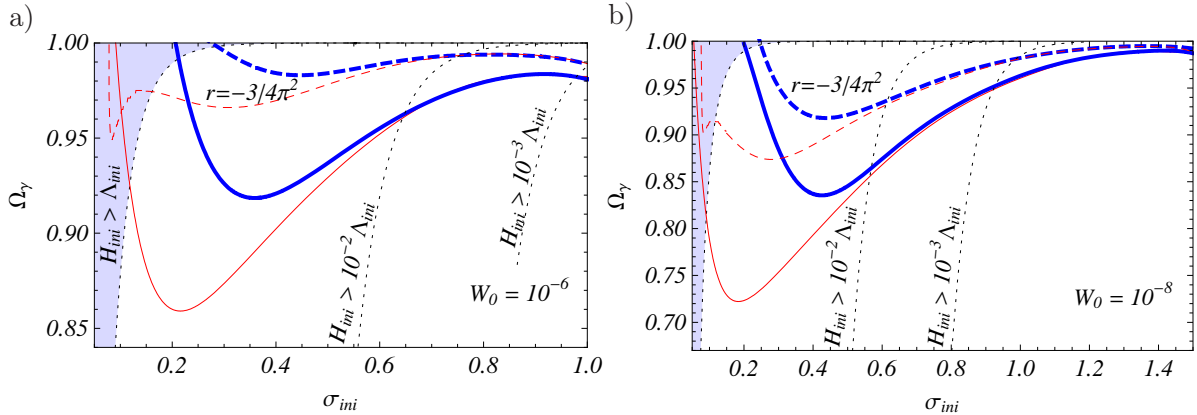

  \begin{minipage}{0.47\linewidth}
    a)\\
    {\centering
    \includegraphics[scale=0.75]{Tplot10-6.eps} 
    }
  \end{minipage}
  \hspace{1mm}
  \begin{minipage}{0.47\linewidth}
    b)\\
    {
    \centering
    \includegraphics[scale=0.74]{Tplot10-8.eps}
    }
  \end{minipage}
  \caption{The regions enclosed by the curves correspond to initial conditions leading to modulus
    stabilization in the KKLT model with $c=1/7$ and a) $W_0=10^{-6}$ and b) $W_0=10^{-8}$ as a function
    of the initial abundance $\Omega_\gamma$. The thin/red curves are obtained in absence of the field
    $u$ for $r=0$ (solid) and $r=-3/4\pi^2$ (dashed). The thick/blue curves are obtained for $u$
    integrated in and initially at its minimum $u_{\rm min}$ for $r=0$ (solid) and $r=-3/4\pi^2$ (dashed).
    Black dotted lines correspond to $(H/\Lambda)_{\rm ini}=1,10^{-2},10^{-3}$.}
  \label{fig:plotuT}
\end{figure}

We explore also how the stabilization of the modulus depends on the initial radiation abundance
$\Omega_\gamma$. In Fig.~\ref{fig:plotuT}, we present our results including the condensate dynamics
(blue/thick curves). In the left (right) panel we have set $W_0=10^{-6}$ ($10^{-8}$). The area over the
solid (blue/thick) curve is the set of stabilization-compatible initial conditions at zero temperature.
The area over the dashed (blue/thick) curve describes stabilization when thermal effects are included,
with $r=-3/4\pi^2$. As expected, for low initial radiation abundance, the modulus runs away before the
temperature decreases below $T_{\rm crit}$ and, therefore, stabilization is not possible. This sets a
bound on the initial value of $\Omega_\gamma$, which in our examples is $\Omega_{\gamma,{\rm
    min}}\approx0.98$ ($0.92$) for $W_0=10^{-6}$ ($10^{-8}$) with $r=-3/4\pi^2$.

We contrast these findings with those results obtained in the traditional way, i.e.  in absence of the
dynamical condensate field $u$ (red/thin curves).  Note that the thin and the thick curves only coincide
in a narrow region close to the stabilization point of $\sigma$ ($\sigma_{\rm min}\approx1.22$ for
$W_0=10^{-6}$ and $\sigma_{\rm min}\approx1.68$ for $W_0=10^{-8}$).  This implies that, integrating out
the condensate is only valid for these points, that is, for initial conditions such that the Hubble rate
$H_{\rm ini}$ is between two and three orders of magnitude lower than the condensation scale
$\Lambda_{\rm ini}$ at the beginning of modulus evolution This observation is rather model independent
and has to be taken into account in general when dealing with moduli stabilization driven by gaugino
condensate in an evolving universe.

\section{Conclusions}
\label{sec:conclusions}

We have studied the cosmological evolution of a modulus coupled to a gaugino condensate and its
consequences for moduli stabilization.  Using the KKLT model as an illustrative example, we have shown
that, in presence of background radiation, most of the region yielding modulus stabilization corresponds
to points in parameter space for which the Hubble parameter $H_{\rm ini}$ is close to the scale at which
the gauginos condense $\Lambda_{\rm ini}$. Once the dynamics of the gaugino pairs is included, we have
shown that:
\begin{itemize}
\item integrating out the condensate directly in the superpotential and the K\"ahler potential fails to
  describe moduli dynamics unless $H_{\rm ini}$ is about two orders of magnitude smaller than
  $\Lambda_{\rm ini}$,
\item the dynamics of the condensate sets a stronger bound on the initial conditions leading to
  stabilization.
\end{itemize}
The former point implies that if one is to consider that the condensate has been integrated out before
the modulus starts its evolution, then only the vicinity of the minimum leads to stabilization.

We have also discussed the role of thermal corrections and shown that they do not affect the conclusions
above. Thermal effects impact on the initial friction needed to damp the modulus and stabilize it,
constraining further the range of initial conditions that lead to stabilization.

To conclude, we point out that our observations are model independent and have to be taken into account
in general when dealing with moduli stabilization driven by gaugino condensate in an evolving universe.
Also, our results are expected to hold independently of the choice of parameters. In particular, they
should be valid in settings with a realistic supersymmetry breaking scale.

\section*{Acknowledgments}
We are grateful to Emilian Dudas and Oleg Lebedev for valuable discussions.

\appendix
\section{Equations of motion}
\label{app:eom}

The equations of motion~\eqref{eomsugra}
\be
\nonumber
\ddot \phi^i \, + \, 3 H \dot \phi^i \, 
+ \, \Gamma^i_{j \, k} \dot \phi^j \dot \phi^k \, 
+ \, K^{i \, \bar\jmath} \dd_{\bar\jmath} V \ = \ 0 \, ,
\ee
can be separated into real and imaginary parts (resp. denoted $\phi_r$ and $\phi_i$) and read
\beq
&&\ddot  \phi_r^k \, + \, 3 H \dot \phi_r^k \, 
+ \, \Gamma^k_{l \, m} \rl \dot \phi_r^l \dot \phi_r^m - \dot \phi_i^l \dot \phi_i^m \rr \, 
+ \, \frac{1}{2} K^{k \, \bar l} \frac{\dd V}{\dd \phi_r^l} \ = \ 0 \quad , \nonumber \\
&&\ddot  \phi_i^k \, + \, 3 H \dot \phi_i^k \, 
+ \, \Gamma^k_{l \, m} \rl \dot \phi_i^l \dot \phi_r^m + \dot \phi_r^l \dot \phi_i^m \rr \, 
+ \, \frac{1}{2} K^{k \, \bar l} \frac{\dd V}{\dd \phi_i^l} \ = \ 0 \quad . \label{eomrealim}
\eeq

Applied to the KKLT system~\eqref{KKLT}, we have
\beq
&&\ddot{\sigma_r} + 3 H \dot{\sigma_r} - \frac{1}{\sigma_r} \rl \dot{\sigma_r}^2 
  - \dot{\sigma_i}^2 \rr + \frac{2 \sigma_r^2}{3} \frac{\dd V}{\dd \sigma_r} \ = \ 0 \quad , \nonumber \\
&&\ddot{\sigma_i} + 3 H \dot{\sigma_i} - \frac{2}{\sigma_r} \dot{\sigma_r} \dot{\sigma_i} 
  + \frac{2 \sigma_r^2}{3} \frac{\dd V}{\dd \sigma_i} \ = \ 0 \quad . \label{eomsigma}
\eeq
In section~\ref{sec:full}, we aim at comparing the region of stabilization with the results obtained
in~\cite{Barreiro:2005ua}. However, from then on, we concentrate on the real parts of $\sigma$ and $u$.
Dynamics of the imaginary parts shall be treated elsewhere.

For the exact system~\eqref{VY}, the first equation in~\eqref{eomrealim} becomes
\beq
\ddot{\sigma_r} + 3 H \dot{\sigma_r} - \frac{2}{2\sigma_r - u_r^2 /3} \rl \dot{\sigma_r}^2 
  - \frac{u_r}{3}\dot{u_r}\dot{\sigma_r} \rr 
  + \frac{2 \sigma_r - u_r^2 /3}{6} \rl 2 \sigma_r \dd_{\sigma_r} V 
  + u_r \dd_{u_r} V \rr & =& 0 \,,\nonumber \\
\ddot{u_r} + 3 H \dot{u_r} - \frac{2}{2\sigma_r - u_r^2 /3} \rl \frac{u_r}{3} \dot{u_r}^2 
  - \dot{u_r}\dot{\sigma_r} \rr + \frac{2 \sigma_r - u_r^2 /3}{6} \rl u_r \dd_{\sigma_r} V 
  + 3 \dd_{u_r} V \rr & =& 0 \,.\label{eomsigmaandu}
\eeq

In general, it proves much easier to re-express eqs.~\eqref{eomsugra} in terms of the canonical momenta
associated to the fields $\Pi_i  = {\dd {\mathcal L}_{\rm kin}}/{\dd \dot{\phi^i}}$
\beq
\phi^{\, \prime \, i}  &=&  \frac{1}{H}\, \dot \phi^i \rl \Pi_i \rr \, , 
\label{eommomenta} \\
\Pi^{\, \prime}_i &=& - \, 3 \Pi_i + \frac{1}{H} \, \frac{\dd}{\dd \phi^i} 
\, \rl {\mathcal L}_{\rm kin} - V ( \phi^i ) \,  \rr \, , \nonumber
\eeq
where prime denotes derivative with respect to $N$ (i.e $\dot{x} = H
x'$). These equations can then be integrated numerically in order to obtain the solutions $\phi^i(N)$.

\addcontentsline{toc}{chapter}{Bibliography}

\providecommand{\bysame}{\leavevmode\hbox to3em{\hrulefill}\thinspace}

\end{document}